\documentclass[a4paper]{article}

\usepackage{icrc2013}

\newcommand{\avg}[1]{\left< #1 \right>} 

\title{A novel LIDAR-based Atmospheric Calibration Method for Improving the Data Analysis of MAGIC}

\shorttitle{MAGIC LIDAR Atm. Cal.}

\authors{
Christian Fruck$^{1}$,
Markus Gaug$^{2,3}$,
Roberta Zanin$^{4}$,
Daniela Dorner$^{5}$,
Daniel Garrido$^{2,3}$,
Razmik Mirzoyan$^{1}$,
Lluis Font$^{2,3}$
for the MAGIC Collaboration.
}

\afiliations{
$^1$ Max-Planck-Institut f\"ur Physik, M\"unchen, Germany \\
$^2$ F{\'i}sica de les Radiacions, Departament de F{\'i}sica, Universitat Aut{\`o}noma de Barcelona, 08193 Bellaterra, Spain \\
$^3$ CERES, Universitat Aut{\`o}noma de Barcelona-IEEC, 08193 Bellaterra, Spain \\
$^4$ Universitat de Barcelona, 08014 Barcelona, Spain \\
$^5$ Universit\"at W\"urzburg, 97070 W\"urzburg, Germany \\
}

\email{fruck@mpp.mpg.de}

\abstract{
A new method for analyzing the returns of the custom-made 'micro'-LIDAR system, which is operated along with the two MAGIC telescopes, allows to apply atmospheric corrections in the MAGIC data analysis chain. Such corrections make it possible to extend the effective observation time of MAGIC under adverse atmospheric conditions and reduce the systematic errors of energy and flux in the data analysis.\\
LIDAR provides a range-resolved atmospheric backscatter profile from which the extinction of Cherenkov light from air shower events can be estimated. Knowledge of the extinction can allow to reconstruct the true image parameters, including energy and flux. Our final goal is to recover the source-intrinsic energy spectrum also for data affected by atmospheric extinction from aerosol layers, such as clouds.
}


\begin{document}
\maketitle

\section{The impact of atmospheric conditions on IACT observations}

For the analysis of data from Imaging Air-shower Cherenkov Telescopes (IACTs), precise knowledge of the state of the atmosphere during observations is of great importance. Up to now, data taken under non-optimal conditions had to be rejected in order not to introduce biases in the energy and flux reconstruction. For the MAGIC collaboration, we are for the first time developing a technique for correcting the effect of a variable atmospheric transmission in the data analysis. The most important input for this correction comes from a low-power elastic LIDAR that we operate together with our telescopes and which provides us with real time, range-resolved information about the variable atmospheric conditions in our field of view.

\section{The MAGIC 'micro' LIDAR system}

\begin{figure}[h]
  \centering
  \includegraphics[width=0.5\textwidth]{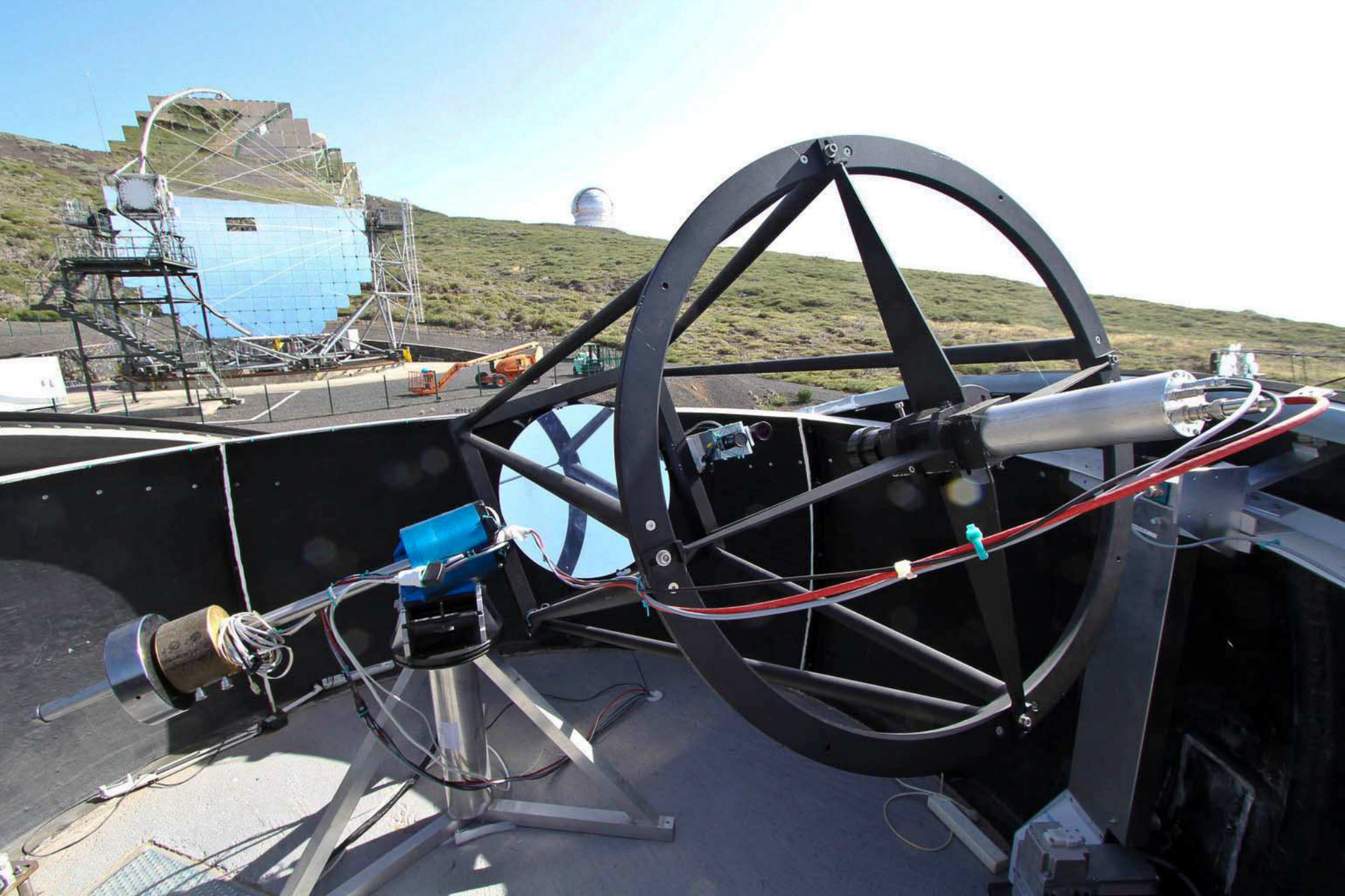}
  \caption{The MAGIC 'micro' LIDAR system on top of the control house with the MAGIC II telescope and GRANTECAN in the background (image credit: Robert Wagner).}
	\label{fig:lidar_sketch}
\end{figure}

\begin{figure}[h]
  \centering
  \includegraphics[width=0.5\textwidth]{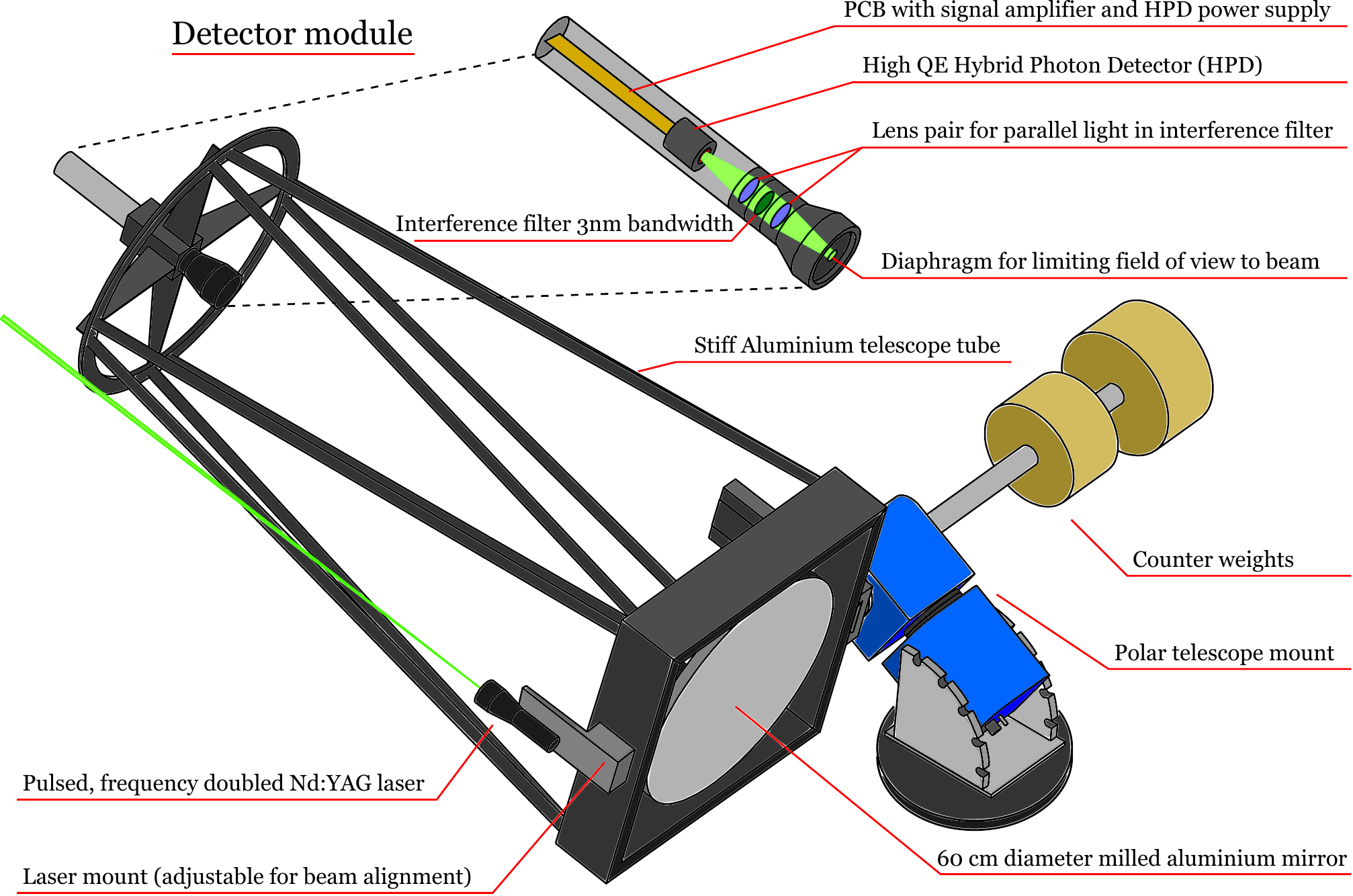}
  \caption{Sketch of the MAGIC LIDAR system with detailed description of all its components.}
	\label{fig:lidar}
\end{figure}

\begin{figure}[h]
  \centering
  \includegraphics[width=0.485\textwidth]{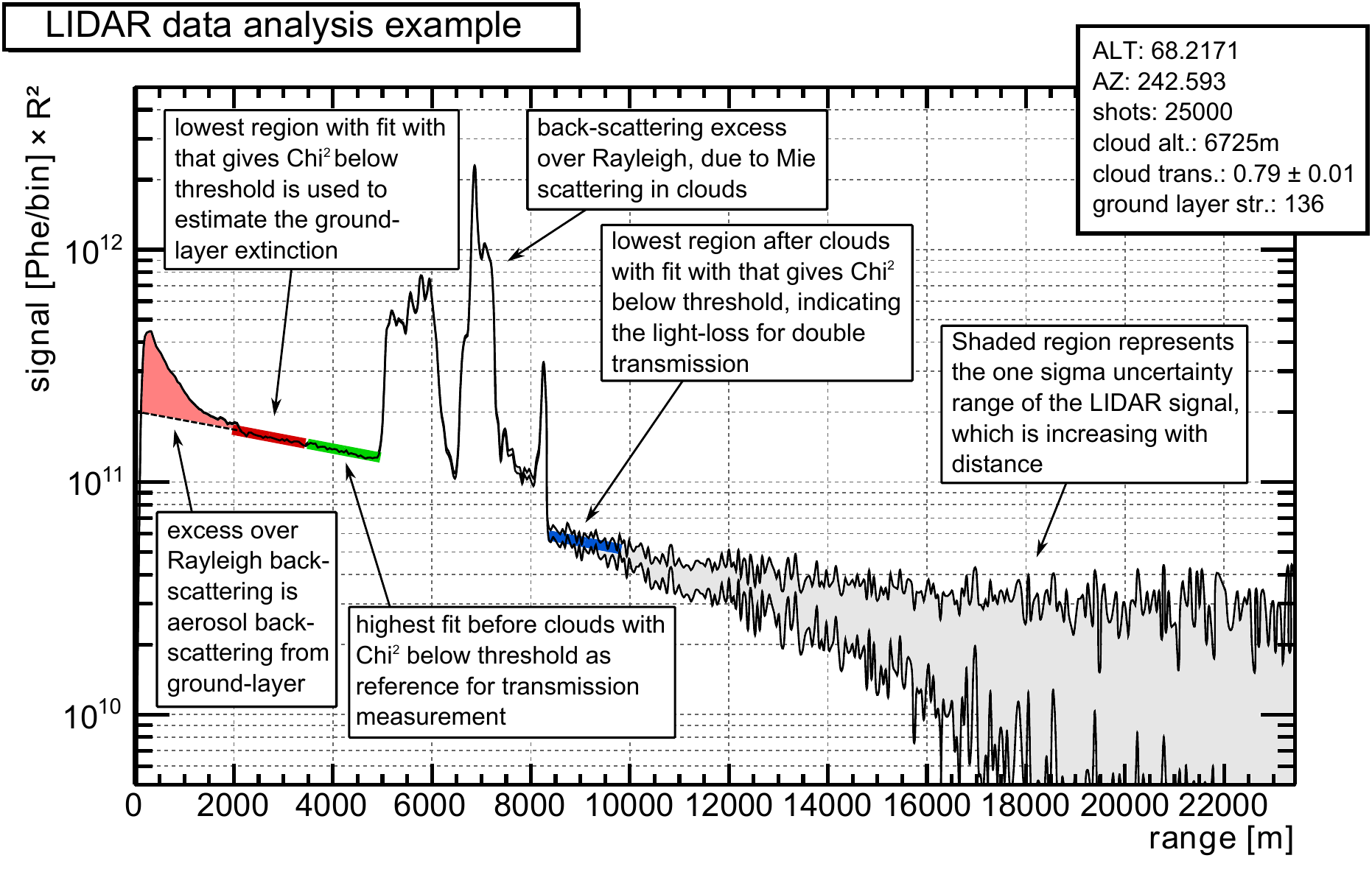}
  \caption{Illustration explaining the data analysis algorithm. The plot is showing a real data example of a range-corrected LIDAR return signal. The multiplication by the square of the distance of the scattering region $R^2$ is done to remove the dominant dependence of the detector solid angle collecting the scattered light.}
	\label{fig:algorithm}
\end{figure}

\begin{figure}[h]
  \centering
  \includegraphics[width=0.48\textwidth]{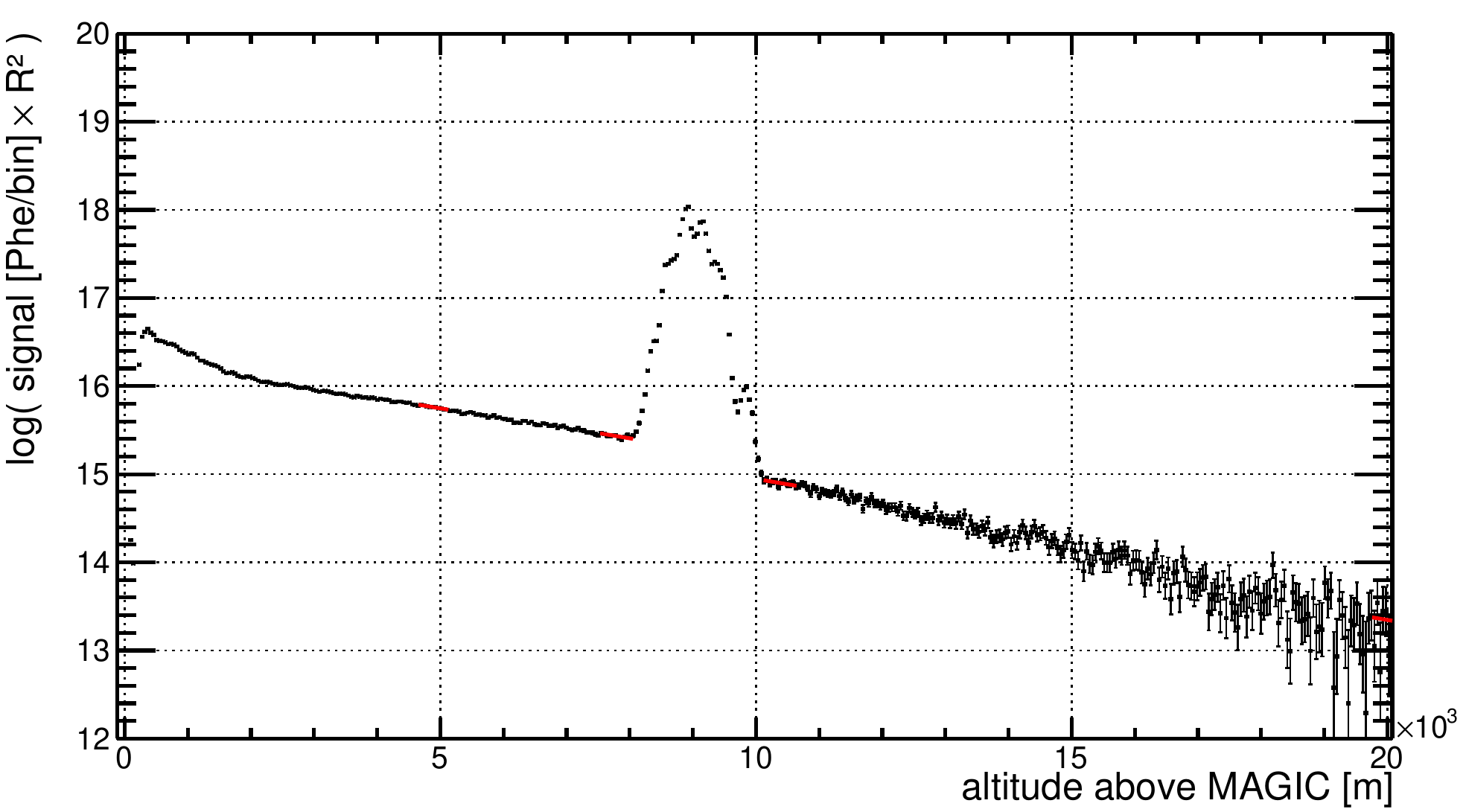}
  \includegraphics[width=0.48\textwidth]{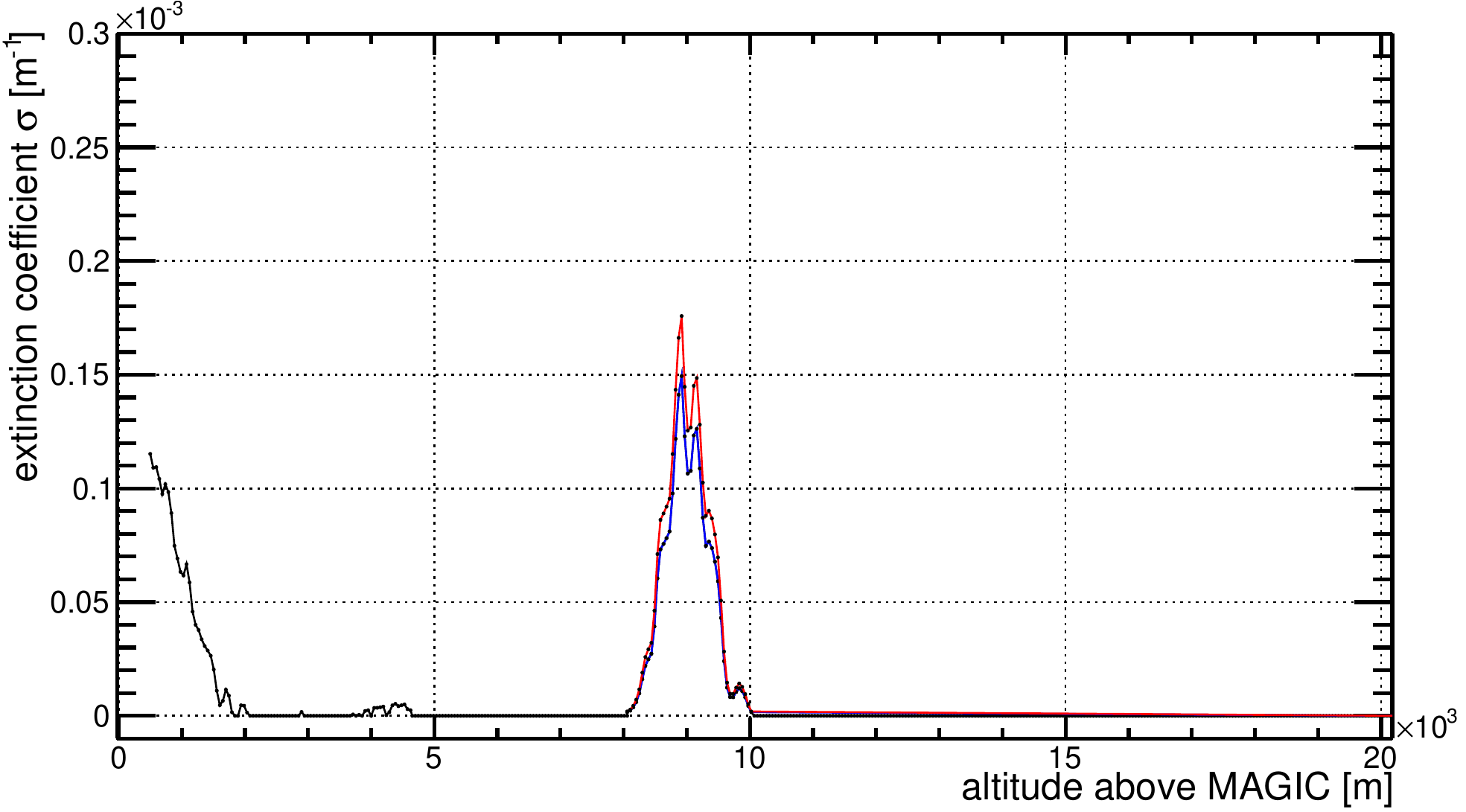}
	\includegraphics[width=0.48\textwidth]{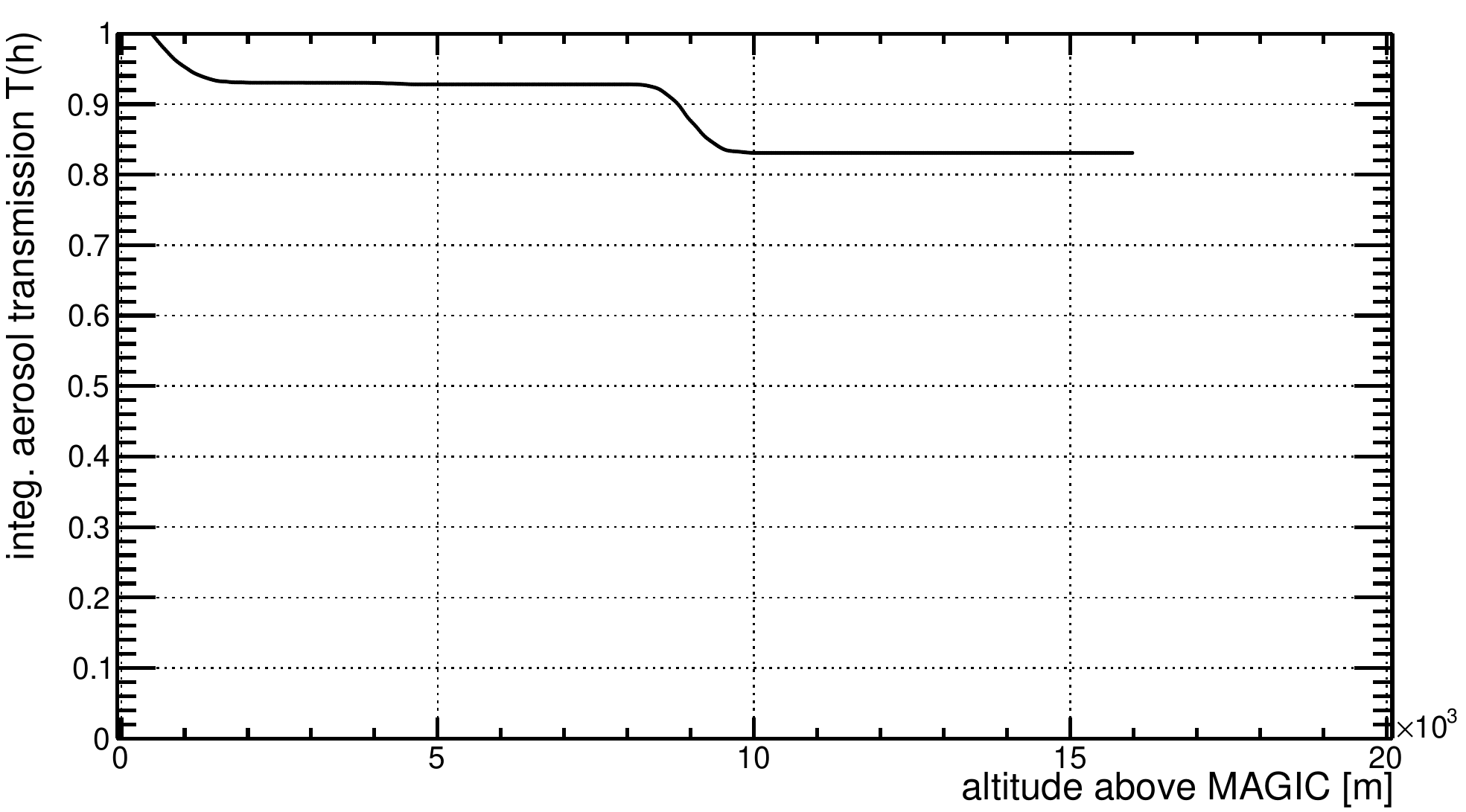}
  \caption{The analysis algorithm for analyzing LIDAR data: range ($R$) corrected signal (photon counts $\times R^2$, top), total aerosol volume extinction coefficient $\sigma_a(h)$ determined with the two different methods (center, blue: cloud transmission method, red: fixed LIDAR-ratio) and the integral atmospheric aerosol transmission $T(h)$ (bottom).}
	\label{fig:data}
\end{figure}

The LIDAR system (fig. \ref{fig:lidar_sketch}) that is operated together with the MAGIC telescopes is a single-wavelength elastic Rayleigh LIDAR operating at 532nm wavelength. This wavelength is not too far from where the Cherenkov spectrum is peaked ($\sim$330nm) and it provides a ratio between cloud/aerosol scattering and molecular scattering that is close to unity in typical cases. Its pulse energy is 5$\mu$J and the repetition rate is 1kHz. The low output power of only 5mW reduces the impact on observations by other telescopes and MAGIC itself to a minimum. The backscattered light is collected by a 60cm diameter mirror with 150cm focal length and focused on the detector module. The detector module consists of a diaphragm of 6mm diameter, a pair of lenses with an interference filter in-between and a Hybrid Photo Detector (HPD) with excellent single-photon detection efficiency as well as charge collection capabilities. The interference filter is used to select a wavelength band of 3nm width around the laser wavelength. The HPD is a Hamamatsu R9792U-40 that provides a peak quantum efficiency of 55$\%$ and excellent charge resolution. An overview of all the hardware components is given in figure \ref{fig:lidar} \cite{bib:schwarz, bib:fruck}. \\
The signal is amplified inside the detector module and digitized by an FADC computer card. For one transmission measurement, 50000 single laser shots are collected and analyzed. The returned signal is divided into three different regions: a pre-trigger region for the Light of Night Sky (LoNS) background subtraction, a short range region with signal pile-up requiring charge integration, and a long range region, from where single photo electron (ph.e.) events are counted. This ensures the large dynamic range needed for a signal region ranging from distances of 0.5km to 18km. The LIDAR returns are analyzed with two algorithms, that make use of regions with a dominant Rayleigh scattering component before and after cloud/aerosol layers and the excess due to additional scattering in-between (see fig. \ref{fig:algorithm} for an illustration of the algorithms and fig. \ref{fig:data} for a real data example). The first method measures the total attenuation of the cloud layer, by comparing the signal before $S_1$ and after the cloud $S_2$ and using the excess over the Rayleigh scattering part of the signal $ex(h)$ to extrapolate to the total aerosol volume scattering coefficient as a function of height $\sigma_a(h)$.

\begin{equation}
	\sigma_a(h) = 
	\frac{ \sqrt{\frac{S_2}{S_1}} }{ \displaystyle\int_{h_1}^{h_2}{ex(h) \mathrm{d}h } } \cdot ex(h)
\end{equation}

The second method uses an empirically determined LIDAR-ratio of $K = 26.0\pm6.0$ for typical thin clouds over La Palma to calculate $\sigma_a(h)$ directly from the excess $ex(h)$ and the known total molecular scattering coefficient $\sigma_m$. The LIDAR-ratio was determined by using the extinction coefficient calculated with the first method and the backscattering coefficient from the LIDAR signal for a selected sample of clouds. Similar values have already been found by \cite{bib:ackermann}. Both methods are applied in measurements taken under different conditions and can be used for cross-checks in the overlapping region.

\begin{equation}
	\sigma_a(h) =  K \cdot \sigma_m(h) \cdot \frac{ ex(h) }{ S_R(h) }
\end{equation}

The final product of the LIDAR measurements is a vertical profile of the total extinction coefficient $\sigma_a(h)$ of everything that is not due to Rayleigh scattering. This profile can be converted into a cumulative transmission profile $T_a(h)$ for the aerosol component and will serve as input information for all further atmospheric corrections in the MAGIC data analysis chain.

\begin{equation}
	T_a(h) = \int_\mathrm{0}^h{\sigma_a(h) \mathrm{d}h}
	\label{eq:T}
\end{equation}

\section{A simple first order approach to atmospheric corrections for IACTs}

\begin{figure}[h]
  \centering
  \includegraphics[width=0.45\textwidth]{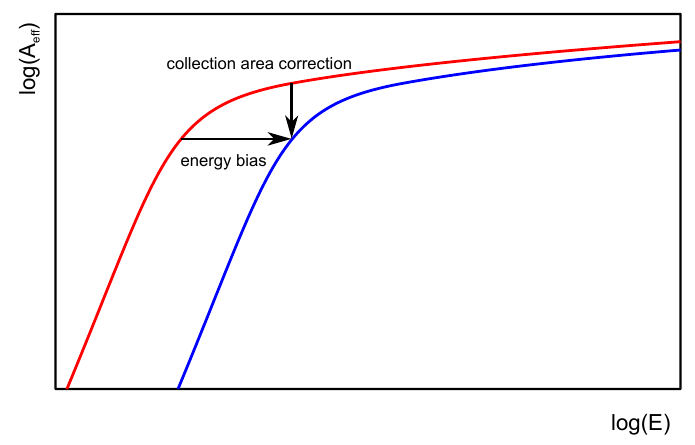}
  \caption{This sketch illustrates, how to do a first order correction to IACT images that are affected by aerosol extinction. The energy has to be up-scaled to correct for the aerosol extinction but the collection area should be evaluated at the apparent (smaller) energy. As a result, the curve that describes the effective collection area $A_\textrm{eff}(E)$ gets simply shifted to the right.}
	\label{fig:area}
\end{figure}

The analysis of IACT data with corrections for variable atmospheric aerosol transmission can be arbitrarily sophisticated, depending on the method used. Whenever tailored Monte Carlo simulations are used, a large variety of simulation sets would be required to reproduce variable conditions with reasonable precision. Other techniques are possible, like the scaling of the light content for each pixel of the IACT camera individually to account for aerosol absorption. With the construction of sophisticated likelihood, which includes atmospheric extinction, such a technique would require stereo information already on image cleaning level, since altitude information would be required for each pixel.\\
For the moment, we use a very simple approach, which works well for low to medium aerosol extinction. The primary parameter that is affected by clouds or aerosols is the light content of the air-shower images in the camera. The shape of the images might be altered as well, if the cloud affects only a part of the air-shower. But this can be considered a second order effect for optically thin clouds. Multiple scattering will be neglected as well in this approach. The change in light intensity has two primary effects, when observed with IACTs. First of all, the energy reconstruction, which mainly depends on the size parameter of the Hillas parametrization of the recorded image, will be biased \cite{bib:hillas, bib:garrido}. The second effect concerns the trigger efficiency, that will decrease, close to the threshold, as well as for higher energies at large impact parameters (see fig.1 in \cite{bib:garrido} for a more comprehensive illustration of that effect). One can now assume, that ``an air-shower that is affected by aerosol extinction looks like an air-shower of smaller energy''. This means that it will be treated, in first order, like an air-shower from a lower energy primary particle, regarding trigger efficiency and energy reconstruction.\\
The correction can be done by scaling the size parameter to account for lower light content in the air-shower image due extinction and to evaluate the effective collection area $A_\textrm{eff}(E)$ at the energy before up-scaling, see figure \ref{fig:area} for explanation.

	\subsection{Correcting the energy}
	
	Correcting the energy is quite straightforward if one has a good approximation of the total light extinction. In such a case, the energy estimation $E_\mathrm{est}$ just has to be up-scaled by one over the weighted aerosol transmission of the atmosphere $\overline{\tau}$.
	
	\begin{equation}
		\overline{\tau} = \int_{0}^{\infty}{\epsilon(h) \cdot T_a(h) \ \mathrm{d} h}
	\end{equation}
	
	Here $\epsilon(h)$ is the normalized estimated light emission profile of those photons of the air-shower which are contained in the camera images and $T_a(h)$ is the integral aerosol transmission from $h$ to the ground (see eq. \ref{eq:T}). In first order and assuming a linear correlation between light yield of an air-shower and the energy of the primary $\gamma$-particle, one can correct the estimated energy $E_\mathrm{est}$ as follows:
	
	\begin{equation}
		E_\mathrm{true} = \frac{E_\mathrm{est}}{\overline{\tau}}
	\end{equation}
	
	In this way, the energy estimation of each event can be corrected using the real-time range-resolved information of the atmospheric aerosol scattering.

	\subsection{Correcting the effective collection area}
	
	The energy correction is quite straightforward. However the correction of the reduced collection area is more elaborate. In principle, one can simply evaluate the corresponding effective collection area from MC-data at the energy before correction $A(E_\mathrm{est})$. One could just re-weight each event by $A(E_\mathrm{true}) / A(E_\mathrm{est})$ to compensate for the events that are not triggered due to the reduced light yield. However, care has to be taken to estimate the statistical uncertainty in each energy bin correctly.\\
Another possibility is to apply a correction to the effective observation time at the moment when the flux is calculated. The instantaneous energy dependent rate $R(E,t)$ can be expressed as follows:
\begin{equation}
	R(E_\mathrm{true},t) = \frac{\mathrm{d} N(E_\mathrm{true})}{\mathrm{d} t}
\end{equation}
Assuming a certain time interval from $0<t<T$, in which the atmospheric conditions are stable, and the energy correction is known, the rate in that time interval can be written as follows:
\begin{equation}
  \avg{R(E_\mathrm{true})} = \frac{ \displaystyle\int_0^T{\frac{\mathrm{d} N(E_\mathrm{true})}{\mathrm{d} t} \ \mathrm{d} t} }{ \displaystyle\int_0^T{\mathrm{d} t} } = \frac{N(E_\mathrm{true})}{T}
\end{equation}
The true differential flux $F(E,t)$ of a source, observed by an instrument with energy and time-dependent effective collection area $A(E,t)$ can be approximated then by:
\begin{equation}
  F(E_\mathrm{true},t) = \frac{\mathrm{d} N(E_\mathrm{true})}{\mathrm{d} t} \cdot \frac{1}{A(E_\mathrm{est},t)}
\end{equation}
We are counting events in absorption corrected energy ($E_\mathrm{true}$), but evaluating the collection area corresponding to the uncorrected energy $E_\textit{est}$ from aerosol-free Monte Carlo simulations. The time average of the flux can be written as follows.
\begin{eqnarray}
  \avg{F(E_\mathrm{true})} &=&
	\frac{ \displaystyle\int_0^T{\frac{\mathrm{d} N(E_\mathrm{true})}{\mathrm{d} t} \cdot \frac{1}{A(E_\mathrm{est},t)} \ \mathrm{d} t} }{ \displaystyle\int_0^T{\mathrm{d} t} } \\
	&=& \frac{ \displaystyle\int_{N(E_\mathrm{true},0)}^{N(E_\mathrm{true},T)}{\frac{\mathrm{d} N(E_\mathrm{true})}{A(E_\mathrm{est}(N))}} }{ T } \\ 
	\rightarrow
	\avg{F_i} &=&
	\frac{ \displaystyle\sum_{j=0}^{N_i}{\frac{1}{A_{i-\delta_j,j}}} }{ T }
\end{eqnarray}
In the last step, the integral becomes a sum over all events $j$ when going to integer values of $N_i$ with $i$ being the energy bin in $true$ energy (after correction) and $\delta_j$ the energy correction bias for each event. This means that the flux contribution of each event has to be computed with an individual collection area $A_{i-\delta_j,j}$ for each event (at the energy evaluated from aerosol-free MC simulations, before atmospheric correction). An effective collection area for the entire time span can be defined in the following way:
\begin{equation}
  \frac{ N_i }{ T \cdot A_{\mathrm{eff},i} } = \avg{F_i} = \frac{ \displaystyle\sum_{j=0}^{N_i}{\frac{1}{A_{i-\delta_j,j}}} }{ T }
\end{equation}
\begin{equation}
	\Rightarrow A_{\mathrm{eff},i} = \frac{ N_i }{ \displaystyle\sum_{j=0}^{N_i}{\frac{1}{A_{i-\delta_j,j}}} }
\end{equation}

\section{Performance checks on Crab data}

\begin{figure}[h]
  \centering
  \includegraphics[width=0.5\textwidth]{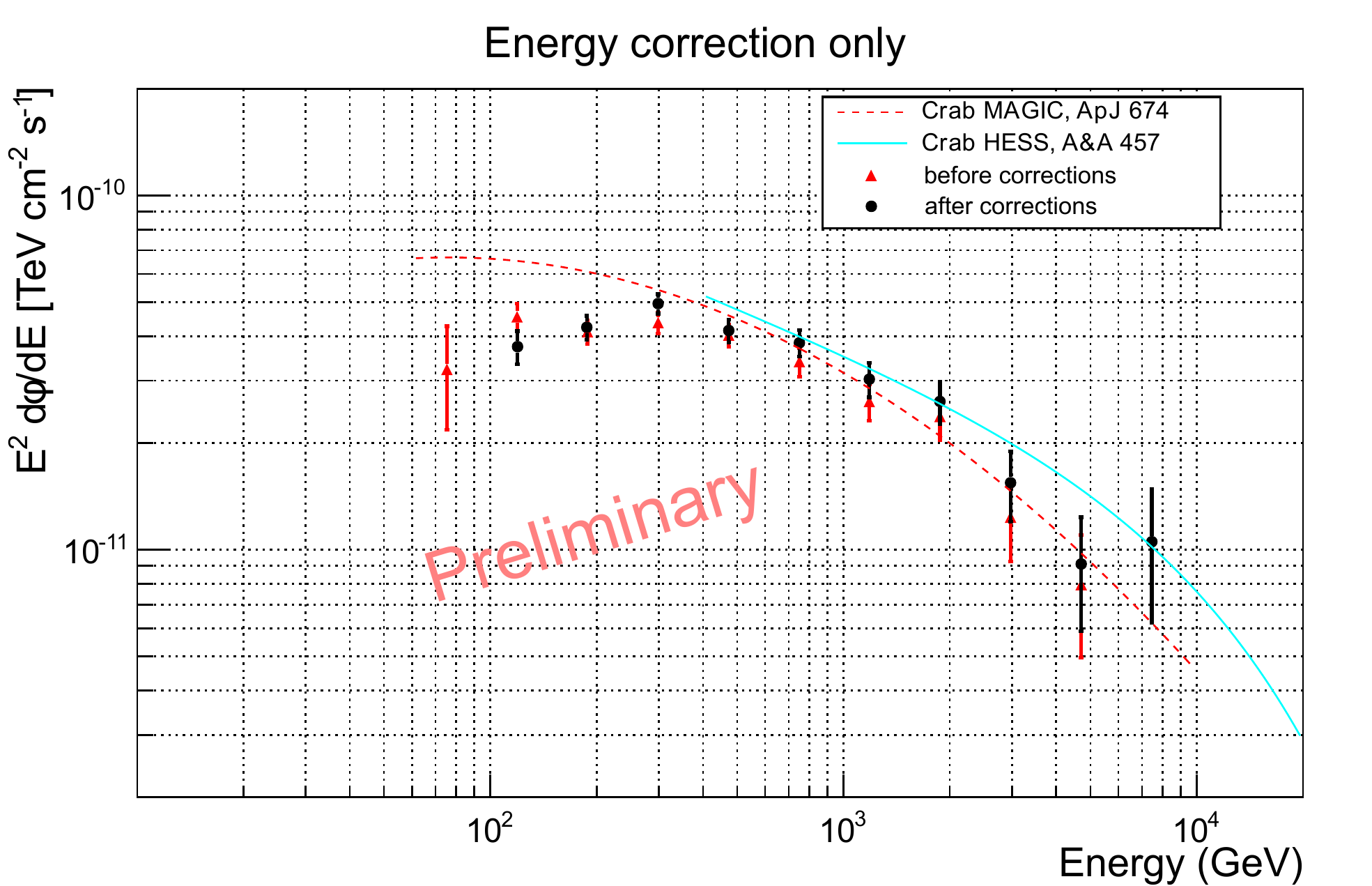}
  \includegraphics[width=0.5\textwidth]{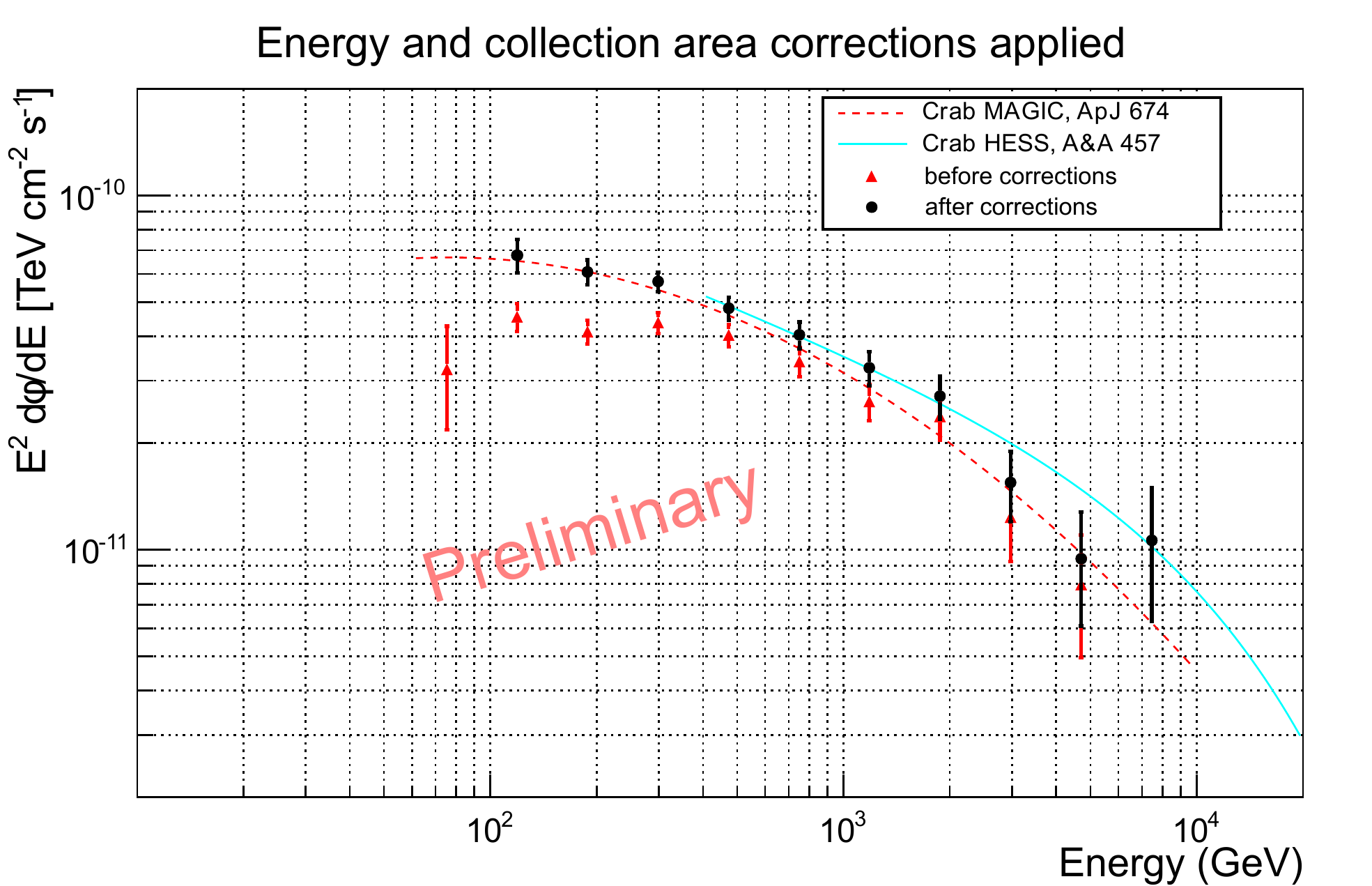}
  \caption{The upper plot shows the Spectral Energy Distribution (SED) of the Crab Nebula without any (red triangles) and after only energy correction (black dots). The lower plot shows the SED from the same data set without any (red triangles) and after energy and colletion area correction (black dots). The data is from 100 minutes of observations during variable moderately cloudy sky conditions ($50-80\%$ aerosol transmission through a cloud layer at 6-8km above the telescopes).}
	\label{fig:corrspec}
\end{figure}

The correction of energy bias and and effective area corrections due to atmospheric extinction are for the first time applied in the MAGIC data analysis software Mars \cite{bib:moralejo}.  The upper plot in figure \ref{fig:corrspec} shows two Spectral Energy Distributions (SEDs) of Crab Nebula data (100 min. observation time) moderate cloudy sky conditions in two days of February 2013. The red SED does not contain any corrections for the energy bias due to aerosol scattering, while the one in black was treated as described above for correcting the energy bias. The lower plot shows the same data set, with also corrections for the collection area applied. It can be seen, that the SED for this source is fully recovered to match the published curves. The lowest energy data point is lost because events migrated to higher energies. But the migrating events allow to but one point more at higher energy.

\section{Conclusions}
Using the optimized atmospheric calibration technique in the MAGIC data analysis chain should enable a reliable use of data taken during moderate cloudy conditions. This way, the effective duty cycle of the telescopes will be extended by up to 15$\%$. Some low energy events close to the threshold will be lost, because they do not trigger any more. But for the higher energies, MAGIC will gain significantly in observation time.


\end{document}